# Spectroscopic ellipsometry of CsPbCl$_3$ perovskite thin films

Sana Khan[1,2], Piotr J. Cegielski[1], Manuel Runkel[3], Thomas Riedl[3], Maryam Mohammadi[1,*], Max C. Lemme[1,2,*]

[1] AMO GmbH, Otto-Blumenthal-Straße 25, Aachen, 52074, Germany
[2] Chair of Electronic Devices, RWTH Aachen University, Otto-Blumenthal-Straße 25, Aachen, 52074, Germany
[3] Institute of Electronic Devices, University of Wuppertal, Wuppertal, Germany

*Corresponding Authors: max.lemme@eld.rwth-aachen.de, mohammadi@amo.de

Designing optoelectronic devices based on cesium lead chloride (CsPbCl$_3$) perovskites requires accurate values of their optical constants. Unfortunately, experimental data for this material is very limited thus far. Therefore, here, we applied spectroscopic ellipsometry (SE) to measure the complex optical constants of thermally evaporated CsPbCl$_3$ thin films with different thicknesses on Si/SiO$_2$ substrates. The data were corroborated with scanning electron microscopy (SEM) images and absorption spectroscopy. An optical dispersion model was developed to derive the complex optical constants and film thicknesses. The Tauc-Lorentz model, in conjunction with two harmonic oscillators, was used to extract the required parameters. The extinction coefficient spectrum exhibited a sharp absorption edge at 411 nm, consistent with the absorption spectrum. In addition, the optical bandgap of the film was calculated from the absorption spectra and SE data. The experimental values agree well with the simulation results, with values of ~ 2.99 eV for different film thicknesses. This work provides fundamental information for designing and modeling CsPbCl$_3$-based optoelectronic devices.

All-inorganic cesium lead halide perovskites ($CsPbX_3$, X= Cl, Br, I) have generated considerable interest for their potential use in various emerging optoelectronic devices, including solar cells, lasers, light-emitting diodes, and photodetectors.[1–10] $CsPbCl_3$, in particular, is recognized as a potential candidate for ultraviolet (UV) photodetectors, as well as violet/blue light-emitting diodes and lasers, owing to its wide bandgap, excellent response to UV light, and exceptionally bright violet/blue color emission.[11–16] For these applications, the thickness of the perovskite layer impacts device performance.[17,18] This relationship has not been thoroughly explored through electro-optical simulations, which are highly efficient for predicting performance variations based on material properties.

A comprehensive understanding of the complex optical constants is crucial for conducting electro-optical simulations. Complex optical constants, *i.e.,* the refractive index $n$ ($\lambda$) and extinction coefficient $k$ ($\lambda$), are vital for representing the materials in device models. Spectroscopic ellipsometry (SE) is a well-established, non-destructive technique that accurately measures material properties, including the refractive index, extinction coefficient, bandgap, and absorption coefficient.[19] The complex optical constants and thicknesses of thin films can be measured by analyzing changes in the polarization state of light reflected from the surface of the films.[20] Nevertheless, interpreting SE results is challenging, as the measured quantities need to be linked to actual material properties. SE relies heavily on mathematical models, which must be developed to extract the optical constants of the material. Extensive research has been conducted on determining optical constants for organic-inorganic perovskites.[21–24] All-inorganic perovskites, such as $CsPbI_3$ and $CsPbBr_3$, have also been investigated.[25–27] However, the optical constants of wide-bandgap $CsPbCl_3$ perovskites have yet to be better understood in order to develop new optoelectronic device applications.

Here, we report the complex optical constants of $CsPbCl_3$ thin films extracted from SE. Owing to the material's extremely poor precursor solubility, depositing a fully covered thin film via conventional solution processing methods is challenging. To overcome this, we utilized thermal evaporation as an effective method for producing high-quality $CsPbCl_3$ thin films with complete coverage following an approach we have published earlier.[28,29] $CsPbCl_3$ thin films of different thicknesses were deposited. An optical model was developed based on the SE results and used to derive the film thickness, as well as the *n* and *k* values. Additionally, the optical bandgaps of the thin films were determined from the absorption results (experimental) and SE results (simulated). The bandgap values determined from both methods are closely aligned with each other and the literature[30], validating the accuracy of our optical model.

Two samples with different film thicknesses, sample A (90 nm) and sample B (180 nm), were prepared via vacuum evaporation, as described in previous work.[28,29] Individual layers of 10 nm CsCl and 5 nm $PbCl_2$ were alternately deposited to produce $CsPbCl_3$ thin films on $Si/SiO_2$ and glass substrates. The perovskite thin films were initially characterized by scanning electron microscopy (SEM) and ultraviolet-visible (UV-Vis) spectroscopy. SEM measurements were performed on a Zeiss SUPRA 60 at 4 kV with a working distance of 3.5 mm. UV-Vis spectroscopy of the $CsPbCl_3$ thin films on glass substrates was performed via a PerkinElmer UV-Vis spectrophotometer. The spectroscopic ellipsometry of the $CsPbCl_3$ thin films was conducted via a J. A. Woollam M-2000 spectroscopic ellipsometer. The acquired data were analyzed and modeled via the J. A. Woollam CompleteEASE software package.

An SEM image of a $CsPbCl_3$ thin film produced via the thermal evaporation method is shown in Figure 1 (a). The corresponding UV-Vis absorption spectra show the characteristic absorption peak of the $CsPbCl_3$ thin film at 407 nm (Figure 1(b)).

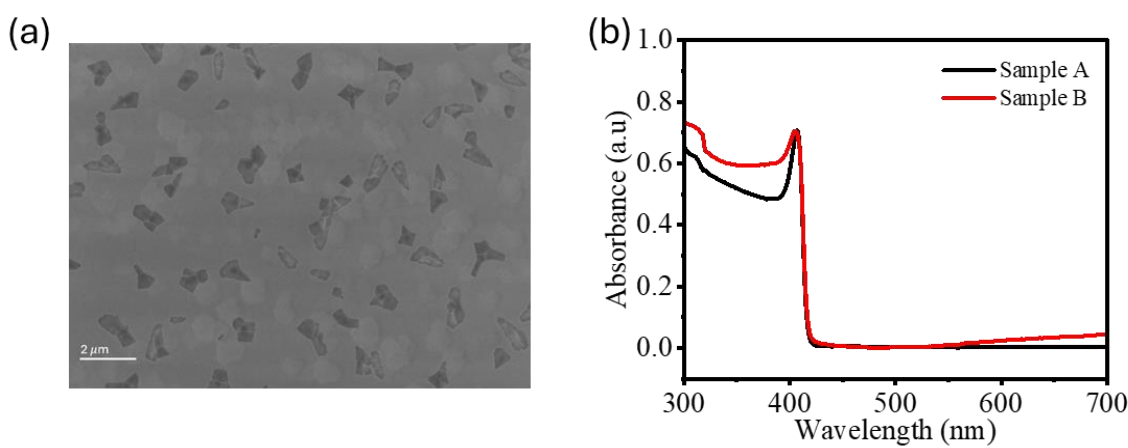

FIG. 1. (a) SEM image of a $CsPbCl_3$ thin film. (b) Normalized UV-Vis absorption spectra of the $CsPbCl_3$

The optical properties of $CsPbCl_3$ thin films were further investigated by SE using a J.A Woollam ellipsometer with a wavelength range from 300 nm to 900 nm. The basic configuration of the spectroscopic ellipsometer is illustrated in Figure 2(a). It comprises a light source and light detector, a stage or sample holder, a polarization state generator, and an analyzer. For a given interface, the ratios of the reflected to incident electric field components for both p-polarized (parallel polarization) and s-polarized (perpendicular polarization) light are known as the Fresnel reflection coefficients. These coefficients depend on the complex refractive indices of both media and the angle at which light strikes the interface. For p-polarized light and s-polarized light, reflection can induce a phase shift. The phase shift, denoted as 'Delta (Δ)', represents the phase difference between the reflected and incident waves. Additionally, reflection affects the amplitude, characterized by 'Psi (Ψ)'. The tangent of Ψ is related to the ratio of the magnitudes of the Fresnel reflection coefficients for the two polarizations, and given by the following relation: [31]

$$\rho = tan\Psi * e^{i\Delta} = \frac{r_p}{r_s} \qquad (1),$$

where $r_p$ and $r_s$ denote the amplitudes of the reflection coefficients for p-polarized light and s-polarized light, respectively. The parameters Δ and Ψ can be determined via SE measurements, which yield both the optical constants and the thickness of the samples.

The propagation of light from a sample can be described via optical models, which can be fitted to the measured SE spectra by adjusting the fitting parameters. A multilayer optical model was built to analyze the SE spectra. It consists of a Si substrate, an interface between the Si substrate and $SiO_2$ layer, a $SiO_2$ layer, a perovskite layer, and a surface roughness layer (Figure 2(b)), since the surface of perovskites is usually rough and must be included in the model for accuracy. This is accomplished by adding a roughness layer atop the solid material, modeled as a mixture of bulk and voids, using the Bruggeman effective medium approximation (EMA). The EMA layer is defined by two key parameters: the thickness and void fraction. Herein, the surface roughness layer is approximated as a 50% mixture of air and the $CsPbCl_3$ film. Each layer of the optical model has several fitting parameters, such as thickness and optical dispersion parameters. The accuracy of the fit is judged by the lowest mean squared error (MSE): [26]

$$MSE = \sqrt{\frac{1}{1p-q}\sum_i^p[(\Psi_{Exp}^i - \Psi_{Cal}^i)^2 + (\Delta_{Exp}^i - \Delta_{Cal}^i)^2]} \times 1000 \quad (2)$$

where p and q are the numbers of measured wavelength points and fitting parameters, respectively, and the subscripts "Exp" and "Cal" refer to the measured and calculated ellipsometry parameters, respectively. As a general rule, an MSE below 20 is typically required to ensure a good match between the model and the actual optical parameters.[32]

Defining the optical response of underlying layers in a multilayer optical model is important for reducing optical correlation effects and obtaining more accurate values of optical constants and thicknesses for the top layer. Therefore, SE spectra of reference Si/SiO$_2$ substrates without CsPbCl$_3$ layers were first measured and analyzed using a three-layer model consisting of a Si substrate, an interfacial layer, and a SiO$_2$ layer.[33]

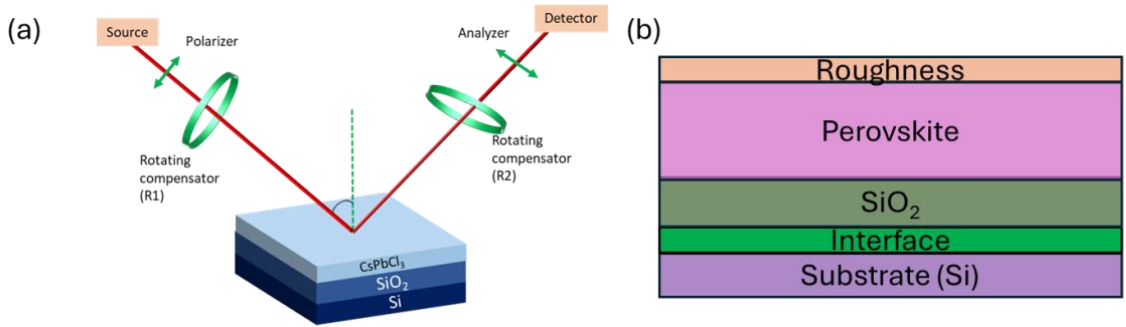

FIG. 2. (a) Schematic illustration of the fundamental structure of the spectroscopic ellipsometer used in the measurement. (b) Schematic of the optical model.

The CsPbCl$_3$ layer was then added atop the SiO$_2$ layer, allowing for precise characterization of its optical properties. The experimental Psi and Delta spectra and the corresponding simulated spectra for samples A and B are shown in Figure 3. To simplify the fitting process, the thickness of the film was estimated via the Cauchy model in the transparent region of CsPbCl$_3$ (> 500 nm). The Cauchy model is specifically suited for estimating the thickness of CsPbCl$_3$ above 500 nm because the material has negligible absorption in this wavelength range, i.e., the extinction coefficient (k) is approximately zero. This allowed for accurate modeling of the refractive index as a function of wavelength via Cauchy's empirical relation: [34]

$$n(\lambda) = A + \frac{B}{\lambda^2} + \frac{C}{\lambda^3} + \cdots \quad (3)$$

where A, B, and C are wavelength-independent material-specific constants. Once the refractive index is well described, the fitting result provides an estimation of the thickness.

The Cauchy layer was then parametrized into a Kramers–Kronig constrained basis spline (B-spline), and the SE data range was extended to the lower wavelength range up to 300 nm. The B-spline layer was used to extract information about the complex dielectric functions of $CsPbCl_3$, as it captures complete details of the dielectric function without prior information about the material's optical behavior. In the final step, the B-spline fitted complex dielectric constants were further parameterized via a general oscillator approach with a Tauc-Lorentz (T-L) oscillator. The complex dielectric function of the energy is defined as:[35]

$$\varepsilon(E) = \varepsilon_1 + i\varepsilon_2 \quad (4)$$

where, $\varepsilon_1$ and $\varepsilon_2$ are the real and imaginary part dielectric function. The T–L oscillator function is given by Equation 5. [34]

$$\varepsilon_2 = \begin{cases} 0 & E \leq E_g \\ \frac{A_n E_{on} Br (E-E_{gn})^2}{(E-E_{on}^2)^2 + Br^2 E^2} \frac{1}{E} & E > E_g \end{cases} \quad n= 1, 2, 3, \ldots \quad (5)$$

The parameters $E_O$, Br, A, and $E_g$ represent the center energy, broadening, amplitude, and band gap per oscillator, respectively, in units of energy (eV).

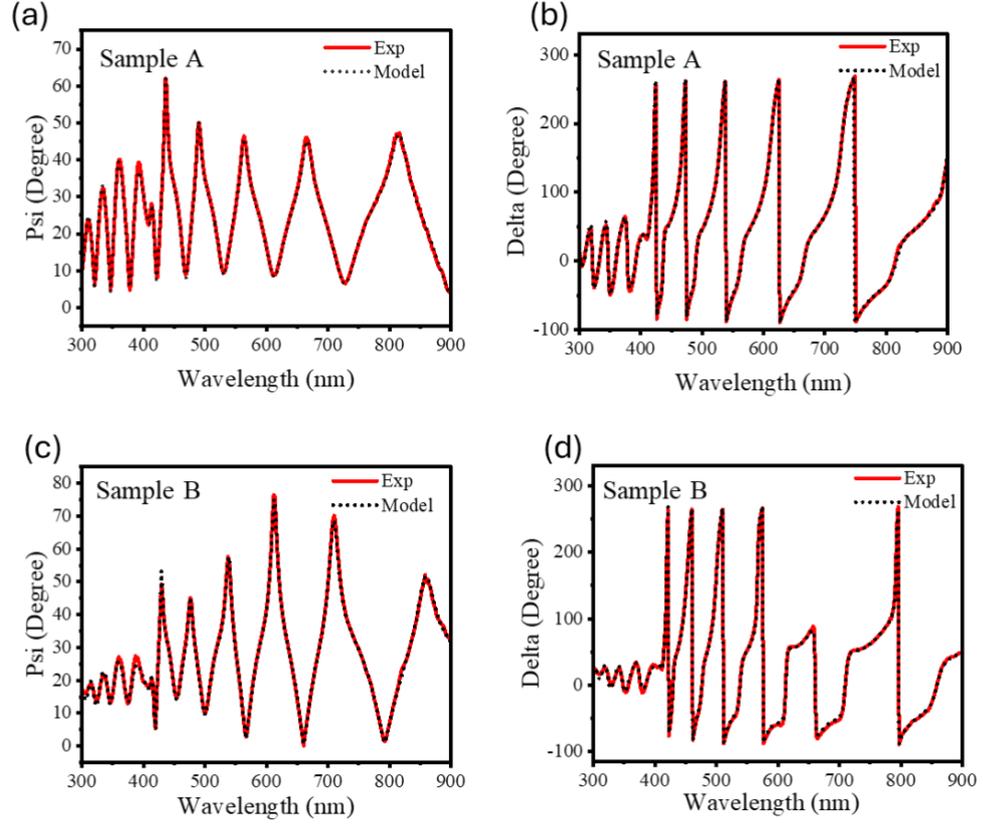

FIG. 3. SE experimental spectra (solid lines) and simulated spectra (dotted lines) of sample A (a and b) and sample B (c and d) measured at a 75° angle.

The fitting parameters of the Tauc-Lorentz oscillator were varied until the best match between the experimental raw data and the simulated data was achieved. Two harmonic oscillators were also used with the Tauc-Lorentz model to better fit the experimental data. The best fit was achieved for thicknesses of approximately 92.73 nm and 187.18 nm for sample A and sample B, respectively. These values are consistent with the expected film thicknesses of 90 nm and 180 nm based on the deposition parameters. The MSE value of the final fit was low, at 3.15 and 6.59 for samples A and B, respectively, demonstrating the reliability of our developed model. More details about the parameters used in the fitting are provided in Tables I and II.

TABLE I. Parameters from Tauc-Lorentz Oscillator.

| Thickness (Deposition parameters) | Thickness (SE) | $E_{inf}$ (eV) | Tauc-Lorentz Oscillator | | | | MSE |
| --- | --- | --- | --- | --- | --- | --- | --- |
| | | | A (eV) | Br (eV) | Eo (eV) | Eg (eV) | |
| 90 nm | 92.73 nm | 0.88 | 581.75 | 0.06 | 3.00 | 2.97 | 3.15 |
| 180 nm | 187.18 nm | 0.59 | 430.43 | 0.091 | 2.99 | 2.95 | 6.59 |

TABLE II. Harmonic Oscillator for sample A and sample B.

| No.of Osc. | A (eV) | Br (eV) | Eo (eV) |
| --- | --- | --- | --- |
| Harmonic Oscillator for sample A | | | |
| 1 | 1.74E-07 | 6.17E-08 | 1.06E-08 |
| 2 | 430.43 | 0.091 | 2.99 |
| Harmonic Oscillator for sample B | | | |
| 1 | 1.15 | 0.41 | 4.0 |
| 2 | 50.86 | 0.00 | 15 |

The refractive index *n* and extinction coefficient *k* for each sample extracted from the optical model best fits are shown in Figures 4(a) and 4(b), respectively. The *n* value indicates the sample's ability to refract incident light, whereas the *k* value reflects absorption loss, scattering loss, and similar effects. *n* (Figure 4(a)) and *k* (Figure 4(b)) exhibit a sharp peak and a steep edge at 411 nm (3.01 eV), respectively. The sample becomes transparent ($k \approx 0$) for higher wavelengths. The distinct peak is the characteristic absorption peak of the $CsPbCl_3$ perovskite.[30]

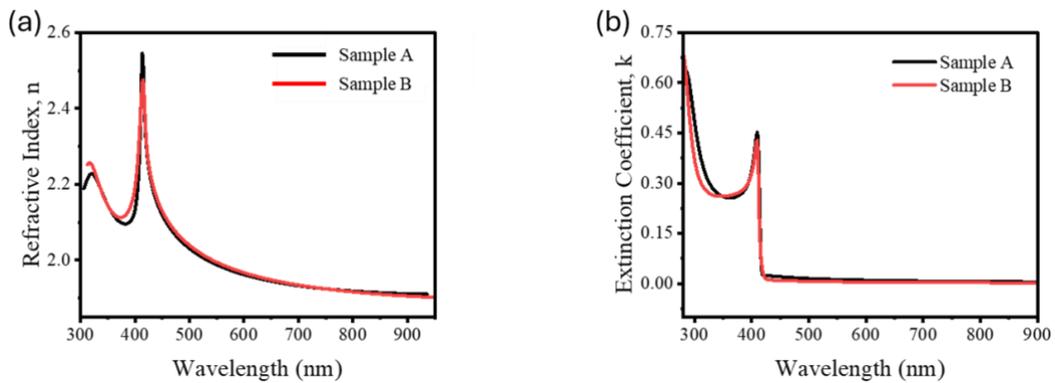

FIG. 4. Optical constants obtained after fitting the optical model to the experimental data. (a) Refractive index *n* and (b) extinction coefficient *k*.

Furthermore, the direct optical band gap of thin films was calculated by linear fitting of the $(\alpha h\nu)^2$ versus $h\nu$ Tauc plot (Figure 5).[30,36] Figures 5(a) and 5(b) show the optical bandgaps derived from the UV-Vis results (experimental data) for samples A and B, respectively. The optical bandgaps obtained for samples A and B are 2.99 eV and 2.99 eV, respectively. To validate the results, the simulation data of the extinction coefficient determined from the optical model were also used to extract the optical bandgap of both samples. Equation 9 was used to calculate the absorption coefficient ($\alpha$) from the extinction coefficient obtained from SE measurements: [25]

$$\alpha = 4\pi k/\lambda \qquad (9).$$

Figures 5c and d show that the bandgap values of 2.99eV derived from the simulated results closely align with the calculated values for samples A and B, respectively. The optical bandgap values also align well with the literature.[30] Based on this comparison, we developed a reliable optical model for evaporated $CsPbCl_3$ thin films, which can be used to design optoelectronic devices.

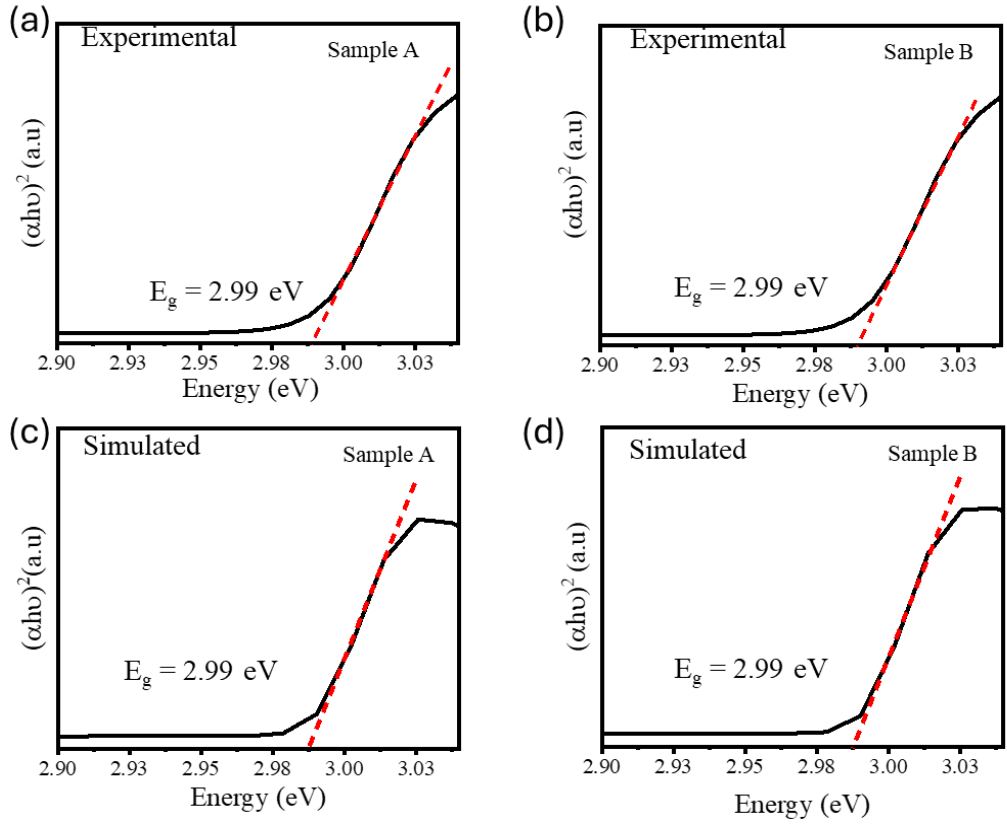

FIG. 5. Plot of $(\alpha h\nu)^2$ vs. energy for samples A and B. Experimental UV-Vis data (a and b) and data simulated from the extinction coefficient obtained by SE fitting (c and d).

In summary, this study presents the optical characterization of thermally evaporated $CsPbCl_3$ thin films. SE measurements were performed on films with thicknesses of 92.73 nm and 187.18 nm to extract the complex optical constants. An optical dispersion model was constructed to evaluate the SE experimental data. The n and k values and the thicknesses of the films were determined by fitting the T-L model in conjunction with two harmonic oscillators. The *k* spectrum exhibited a sharp absorption edge at 411 nm, corresponding to optical bandgaps of 2.99 eV for both samples. These values closely align with those obtained from the UV-Vis results. Our developed optical model contributes to understanding the optical behavior of $CsPbCl_3$ thin films and determining their thickness, thereby facilitating the design of optoelectronic devices.

**Acknowledgments:** This project has received funding from the European Union's Horizon 2020 research and innovation programme under the project FOXES (951774), the German Research Foundation (DFG) through the project Hiper-Lase (GI 1145/4-1, LE 2440/12-1 and RI1551/18-1) and the German Ministry of Education and Research (BMBF) through the project NEPOMUQ (13N17112 and 13N17113).